\documentclass[aps,prl,twocolumn,showpacs,preprintnumbers,amsmath,amssymb,superscriptaddress,bibnotes]{revtex4}

\usepackage{graphicx}
\usepackage{dcolumn}
\usepackage{bm}
\usepackage{color}
\begin{document}


\title{Heat-Capacity Measurements of Energy-Gap Nodes of the Heavy-Fermion Superconductor CeIrIn$_5$ Deep inside the Pressure-Dependent Dome Structure of its Superconducting Phase Diagram }
\author{Xin Lu}

\altaffiliation [X. Lu and H. Lee]{ contribute equally to this work.}

\author{Hanoh Lee}

\altaffiliation[Present address: ]{Department of Applied Physics and Geballe Laboratory for Advanced Materials, Stanford University, Stanford, California 94305, USA}
\affiliation  {Los Alamos National Laboratory, Los Alamos, New Mexico 87545, USA}

\author{T. Park}
\affiliation {Los Alamos National Laboratory, Los Alamos, New Mexico 87545, USA}
\affiliation {Department of Physics, Sungkyunkwan University, Suwon 440-746, Korea}
\author{F. Ronning}
\author{E. D. Bauer}
\author{J. D. Thompson}
\affiliation {Los Alamos National Laboratory, Los Alamos, New Mexico 87545, USA}

\begin{abstract}
We use heat capacity measurements as a function of field rotation to identify the nodal gap structure of CeIrIn$_5$ at pressures to 2.05 GPa, deep inside its superconducting dome. A four-fold oscillation in the heat capacity at 0.3 K is observed for all pressures but with its sign reversed between 1.50 and 0.90 GPa. On the basis of recent theoretical models for the field-angle dependent specific heat, all data, including the sign reversal, imply a d$_{x^2-y^2}$ order parameter with nodes along [110], which constrains theoretical models of the pairing mechanism in CeIrIn$_5$.
\end{abstract}

                            
\maketitle

\begin{figure} [htbp]
\includegraphics[angle=0,width=0.45\textwidth]{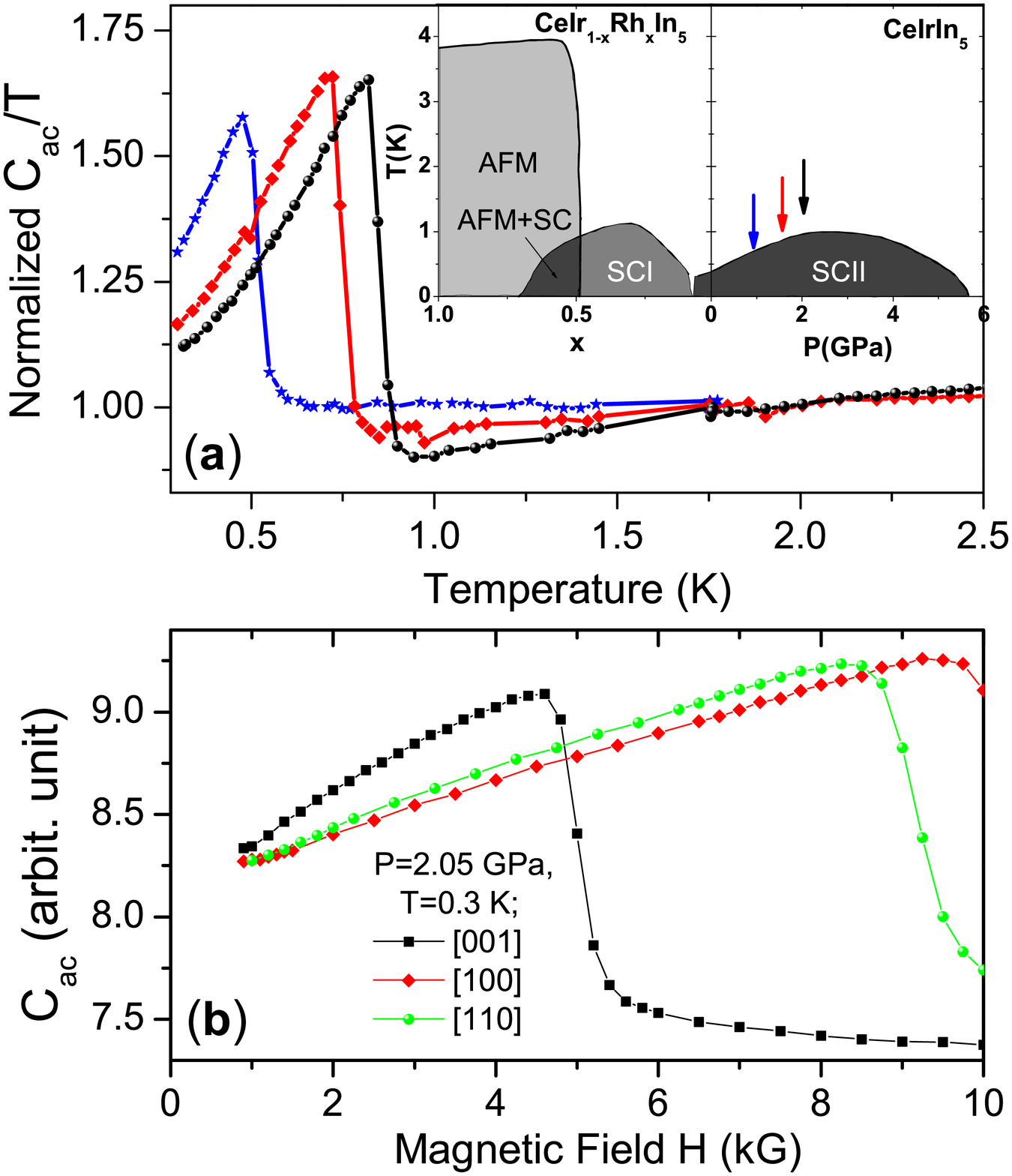}
\vspace{-14pt}
\caption{\label{fig:CeIrIn5phasediagram} (color online). (a) Temperature-dependent heat capacity of CeIrIn$_5$ at pressures P=2.05 (circles), 1.50 (diamonds) and 0.90 GPa (stars) normalized to the values at T=1.8 K. The inset is the phase diagram of Ce(Ir,Rh)In$_5$ as a function of Rh doping and pressure, taken from Ref.\onlinecite{YNakajima08CeIrIn5}; (b) Heat capacity of CeIrIn$_5$ at T=0.3 K and P=2.05 GPa as a function of magnetic field applied along [001] (squares), [100] (diamonds) and [110] (circles) direction, respectively. }
\end{figure}

The Ce-based heavy-fermion family CeMIn$_5$(M=Rh, Co, Ir) has attracted  attention due to their exotic properties and interesting interplay between magnetism and superconductivity \cite{JThompsonReview07}. CeRhIn$_5$ is antiferromagnetic (AFM) at ambient pressure but becomes superconducting as pressure suppresses the AFM order and tunes CeRhIn$_5$ to a quantum-critical point (QCP) \cite{TPark06CeRhIn5}, indicating an intimate relationship between  AFM fluctuations and unconventional superconductivity. CeCoIn$_5$, with the highest superconducting T$_c$=2.3 K among  Ce-based heavy-fermion materials \cite{CPetrovic01CeCoIn5}, is a nodal superconductor with a d$_{x^2-y^2}$ order parameter (OP) that is revealed consistently in a variety of exprimental measurements \cite{WKPark08PRL, KIzawa01CeCoIn5, KAn10CeCoIn5, IEremin08CeCoIn5, CStock08CeCoIn5Neutron}. Cd or Hg doping, acting as an effective negative pressure, tunes the ground state of CeCoIn$_5$ from superconducting to AFM  \cite{LDPham06CeCdCoIn5, EBauerCeCoIn5Hg08}. The resulting temperature-doping phase diagram is very similar to the temperature-pressure diagram of CeRhIn$_5$, revealing a common response of these materials to the interplay between magnetism and superconductivity. CeIrIn$_5$ is also superconducting at ambient pressure with bulk T$_c \sim$ 0.4 K but resistive T$_c \sim$ 1.2 K \cite{CPetrovic01CeIrIn5, ABianchi01CeIrIn5}. This difference in T$_c$s is found as well when magnetic order coexists with superconductivity in CeRhIn$_5$ and doped CeCoIn$_5$ \cite{TParktextureSC}, but there in no long-range AFM in CeIrIn$_5$. Further, the response of CeIrIn$_5$ to effective pressure is different, as shown in the inset of Fig.\ref{fig:CeIrIn5phasediagram}(a). Rh-doping in CeRh$_x$Ir$_{1-x}$In$_5$ acts as an effective negative pressure and induces a dome of superconductivity in proximity to AFM at x$\geq$0.5. T$_c$ goes to zero or approximately zero at x$\sim$ 0.1 before it rises again with applied pressure to form a second dome \cite{SKawasaki06CeIrIn5, MNicklas04CeRhIrIn5}. This diagram is very similar to that of CeCu$_2$(Si$_{1-x}$Ge$_x$)$_2$ as a function of pressure \cite{HQYuan03CeCu2Si2} and has led to the suggestion that the second superconducting dome (SCII), so distant from  AFM order and thus less associated with AFM flutuations \cite{SKawasaki05PRLNQR}, may be due to another pairing mechanism, such as Ce-valence fluctuations \cite{SWatanabe06valence}, and possibly supports a different OP symmetry.

Identifying the nodal gap structure and pairing symmetry in CeIrIn$_5$ is an important step toward resolving the pairing mechanism. Power laws in heat capacity, thermal conductivity \cite{RMovshovich01PRL} and penetration depth \cite{HQYuan09CeIrIn5} are consistent with line nodes in the gap, but these do not probe the node positions, which is needed to determine the allowed OP symmetry. A large anisotropy between in-plane and out-of-plane thermal conductivity measurements on CeIrIn$_5$ has led to a proposed hybrid gap, $k_z(k_x+ik_y)$ \cite{HShakeripour07CeIrIn5,HShakeripour10CeIrIn5}. In contrast, thermal conductivity studies in a magnetic field that is rotated within the basal plane imply a d$_{x^2-y^2}$ gap with line nodes along the [110] direction \cite{YKasahara08CeIrIn5}, similar to CeCoIn$_5$ \cite{KAn10CeCoIn5} and CeRhIn$_5$ under pressure \cite{TPark08CeRhIn5}.  This discrepancy may be due in part to both measurements being made at ambient pressure where the superconductivity of CeIrIn$_5$ sits right at the margin of the two superconducting domes and may be influenced by the competition between residual AFM fluctuations and other pairing interactions, if any. Probing the gap structure in the superconducting state deep inside SCII avoids these complications.   

In this Letter, we report heat capacity measurements of CeIrIn$_5$ under pressures up to 2.05 GPa as a magnetic field is rotated around and through its tetragonal c-axis. A four-fold oscillation of the in-plane heat capacity indicates a gap of d$_{x^2-y^2}$ symmetry with nodes along [110], even deep inside SCII. This OP symmetry should be captured in mechanistic models of superconductivity of CeIrIn$_5$ in SCII.

CeIrIn$_5$ single crystals were grown out of In-rich flux and screened by magnetic susceptibility to ensure the absence of free In. The sample was mounted in a Be-Cu/NiCrAl hybrid clamp-type pressure cell with silicone fluid as the pressure-transmitting medium, which provides a very nearly hydrostatic environment to 25 kbar. The resistive superconducting transition temperature of Sn was measured to determine the pressure at low temperatures. The heat capacity of CeIrIn$_5$ C$_{ac}$ under pressure was measured by an ac calorimetric method wherein heating from an ac current generates an oscillation in the sample temperature $\Delta T_{ac}$ that is measured by a field-calibrated Cr/AuFe(0.07 \%) thermocouple through its ac response; the heat capacity is inversely proportional to $\Delta T_{ac}$, i.e., C$_{ac} \propto 1/\Delta T_{ac}$. Magnetic field rotation was provided by a triple-axis vector magnet that accomodates a $^3$He cryostat with the pressure cell inside.

Figure \ref{fig:CeIrIn5phasediagram}(a) shows the temperature dependence of the heat capacity C$_{ac}$ under pressures P=0.9, 1.5 and 2.05 GPa going deeper into SCII with increased bulk T$_c$=0.50, 0.75 and 0.85 K, respectively. The magnetic field dependent heat capacity C$_{ac}$ is shown in Fig. \ref{fig:CeIrIn5phasediagram}(b) for CeIrIn$_5$ at 0.3 K and 2.05 GPa with the field applied along three major crystallographic orientations: [001], [100] and [110]. The upper critical field along the c-axis H$_{c2}^c$ is smaller than that within the ab plane by a factor of 2, similar to the case of CeCoIn$_5$ \cite{RSettai01CeCOIn5dHva} and CeRhIn$_5$ \cite{TPark08CeRhIn5}. Within the ab plane, H$_{c2}^{[100]}$ along [100] is reproducbily a little larger than H$_{c2}^{[110]}$ in the [110] direction, probably reflecting the d-wave superconducting OP symmetry as discussed later. 

\begin{figure} [htbp]
\includegraphics[angle=0,width=0.45\textwidth]{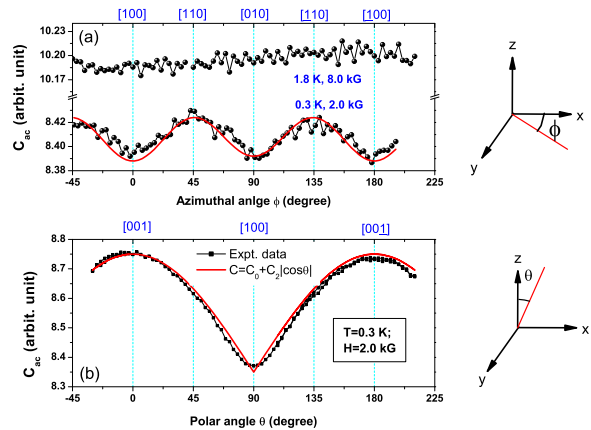}
\vspace{-14pt}
\caption{\label{fig:P2GPaFieldRotation1} (color online). Angle-dependent heat capacity of CeIrIn$_5$ at 2.05 GPa. (a) Azimuthal rotation ($\phi$) of the magnetic field in the ab-plane in the normal (T=1.8 K and H=8.0 kG) and superconducting state (T=0.3 K and H=2.0 kG). The solid line is a fit to a four-fold modulation model $C_{ac}=C_0+C_2cos2(\phi-\phi_0)+C_4cos4\phi$. (b) Polar rotation ($\theta$) of the magnetic field (2.0 kG) in the ac-plane of CeIrIn$_5$ at T=0.3 K. The solid line is a fit to a two-fold oscillation model C$_{ac}$=C$_0$+C$_2|cos\theta|$, where the modulation arises from H$_{c2}$ anisotropy.}
\end{figure}

Figure \ref{fig:P2GPaFieldRotation1} shows the field-angle heat capacity of CeIrIn$_5$ at 2.05 GPa deep inside SCII with magnetic field rotated in the ab (ac) plane. For the field rotated within the ab plane, C$_{ac}$/T is featureless in the normal state at 1.8 K even with magnetic field up to 8.0 kG. An apparent four-fold oscillation is present in $\frac{C_{ac}}{T}(\phi)$ when cooling down to 0.3 K in the superconducting state and applying field at 2.0 kG. A polar sweep at 0.3 K with H=2.0 kG is shown in Fig. \ref{fig:P2GPaFieldRotation1}(b) and a two-fold modulation is observed, which can be fitted to the function $C_{ac}(\theta)=C_0+C_2|cos\theta|$, most likely caused by the H$_{c2}$ anisotropy between c-axis and ab plane.

Theoretical and experimental progress has established field-angle dependent heat capacity (and thermal conductivity) as a powerful probe of nodal locations in the superconducting gap structure  \cite{YMatsuda06fieldrotation,TPark03YNi2B2C,TPark08CeRhIn5,KAn10CeCoIn5}. In the Abrikosov state where supercurrents circulate around the vortex core, extended quasiparticles (QPs) experience a Doppler shift, $\delta E \sim \bf{p_F} \cdot \bf{v_s} $, where \textbf{p$_F$} is the Fermi momentum of QPs and \textbf{v$_s$} is the supercurrent velocity. In nodal superconductors at low temperatures, the Doppler energy shift plays an important role in the QP excitations. Near the nodes, the local superconducting gap is sufficiently small that the Doppler shift  breaks Cooper pairs, \textbf{$\Delta(\bf{p_F})<|\bf{p_F}\cdot \bf{v_s}|$}, and, consequently, the QP density of states (DOS) depends on the field orientation relative to the position of gap nodes. In a classical picture, the QP DOS and thus heat capacity is a minimum for the field along nodal directions and a maximum along antinodal directions. In the case of a d-wave superconductor, a four-fold oscillation emerges in the superconducting state when the field is rotated in the basal plane and the local heat capacity minimum reflects a nodal position on the Fermi surface \cite{IVekhter99}. The four-fold oscillation observed in an azimuthal field sweep for CeIrIn$_5$ at 2.05 GPa implies the presence of vertical line nodes perpendicular to the ab-plane as expected for a gap with d-wave symmetry. A hybrid gap with horizontal line nodes in the equator is inconsistent with the four-fold oscillation observed here.

\begin{figure} [htbp]
\includegraphics[angle=0,width=0.47\textwidth]{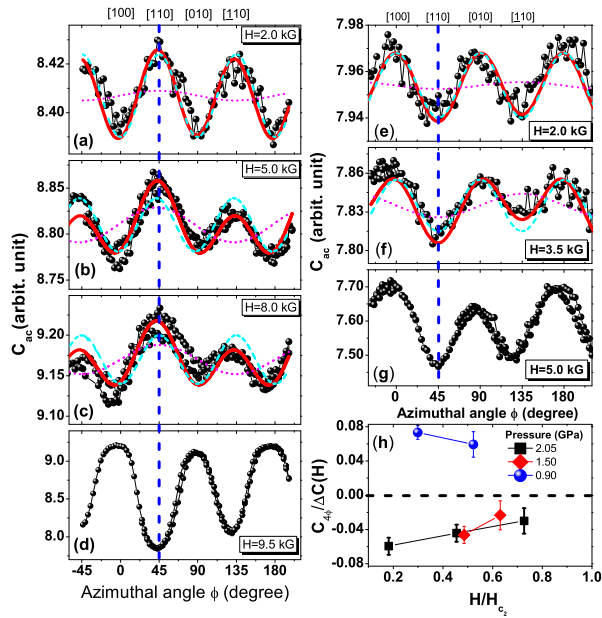}
\vspace{-17pt}
\caption{\label{fig:CeIrIn5FieldRotation} (color online). (a)-(d) Heat capacity of CeIrIn$_5$ at T=0.3 K and P=2.05 GPa as a function of azimuthal field rotation for H=2.0, 5.0, 8.0 and 9.0 kG, respectively. The upper critical field along [100] is 11.0 kG at this temperature and pressure. (e-g) Azimuthal field dependence of the heat capacity of CeIrIn$_5$ at T=0.3 K and P=0.90 GPa  for H=2.0, 3.5 and 5.0 kG, respectively. The upper critical field along [100] is 6.70 kG. Solid lines are a fit to the four-fold oscillation model and the vertical dashed line indicates the [110] direction. The two-fold and four-fold components of the fit in (a)-(c) and (e)-(f) are plotted as dotted and dashed lines, respectively. (h) The four-fold oscillation component C$_{4\phi}$ at T=0.3 K as a function of H/H$_{c2}^{[100]}$ normalized to the corresponding $\Delta C_{ac}(H)=C(H)-C(0)$ at P=2.05 (squares), 1.50 (diamonds) and 0.90 GPa (circles), respectively.}
\end{figure}

The angular location of maxima and minima in C$_{ac}((\phi)$  allows a distinction in the OP symmetry between d$_{xy}$ and d$_{x^2-y^2}$. Figs. \ref{fig:CeIrIn5FieldRotation} (a)-(d) show the heat capacity C$_{ac}$ of CeIrIn$_5$ at 2.05 GPa and 0.3 K as a function of field angle $\phi$ against the [100] direction in the ab plane at different fields up to 9.5 kG. Data at 9.5 kG (Fig. \ref{fig:CeIrIn5FieldRotation} (d)) are qualitatively different from those at lower fields (Figs. \ref{fig:CeIrIn5FieldRotation}(a)-(c)): at 9.5 kG local minima are shifted by 45 degrees and the oscillation amplitude is one order of magnitude larger relative to those in (a)-(c). We note that this magnetic field is very close to H$_{c2}$ for [100] and [110] directions. The in-plane H$_{c2}$ anisotropy will contribute four-fold oscillation in C$_{ac}(\phi)$ with local minina in [110] directions, providing an indirect way to check the crystalline orientation. The offset of C$_{ac}$ between [110] and [$\underline{1}$10] in Fig.\ref{fig:CeIrIn5FieldRotation}(d) indicates a small sample misalignment between the crystal ab-plane and the field xy-plane, which introduces a two-fold oscillation in the field rotation heat capacity. The total heat capacity in magnetic field can thus be writtern as C$_{ac}(\phi)$=C$_0$+C$_2$cos2($\phi-\phi_0$)+C$_4$cos4$\phi$, where C$_0$ is the zero-field heat capacity coming from thermally excited QPs and phonons, the two-fold term is due to the sample misalignment, and the four-fold oscillation C$_4$cos4$\phi$ arises from nodal structure of the superconducting OP. 

Though the local minima of C$_{ac}(\phi)$ are all in the [100] direction and maxima are along [110] for different magnetic fields as shown in Fig. \ref{fig:CeIrIn5FieldRotation}(a)-(c), the naive classical picture would imply a d$_{xy}$ OP symmetry with line nodes along [100] directions for CeIrIn$_5$ at 2.05 GPa. However, Vorontsov and Vekhter have argued that the classical picture holds only when $T/T_c\ll 1$ and $H/H_{c2} \ll 1$, and, consequently, the angular dependence of four-fold oscillations shows a complex evolution across the whole H-T phase diagram \cite{IVekhter07heatcapacity}. In general, anisotropic scattering of QPs due to vortices produced by the magnetic field plays an important role in the DOS variation and thus the heat capacity as a function of the field orientation, which is missed in the classical picture. At an intermediate temperature or moderately high field, angular anisotropy in the heat capacity changes sign and the four-fold oscillation is inverted, showing maxima rather than minima when the field is along the nodal direction. According to their calculation of the field-angle dependent heat capacity of CeCoIn$_5$, where a corrugated cylindrical FS is assumed, our field rotation heat capacity measurments at 0.3 K, with T/T$_c \sim$ 0.35, are in the intermediate region of the H-T phase diagram and a four-fold component with maxima in the nodal directions is expected. Our observation of oscillation maxima in [110] directions for CeIrIn$_5$ at 2.05 GPa indicates that the line nodes are along diagonals of the crystal lattice and favors d$_{x^2-y^2}$ symmetry over d$_{xy}$ for CeIrIn$_5$ at 2.05GPa deep inside SCII. Similar behaviors observed for CeIrIn$_5$ at 1.50 GPa with T/T$_c\sim$0.40 (not shown here) also are consistent with a d$_{x^2-y^2}$ OP symmetry. 

\begin{figure} [htbp]
\includegraphics[angle=0,width=0.45\textwidth]{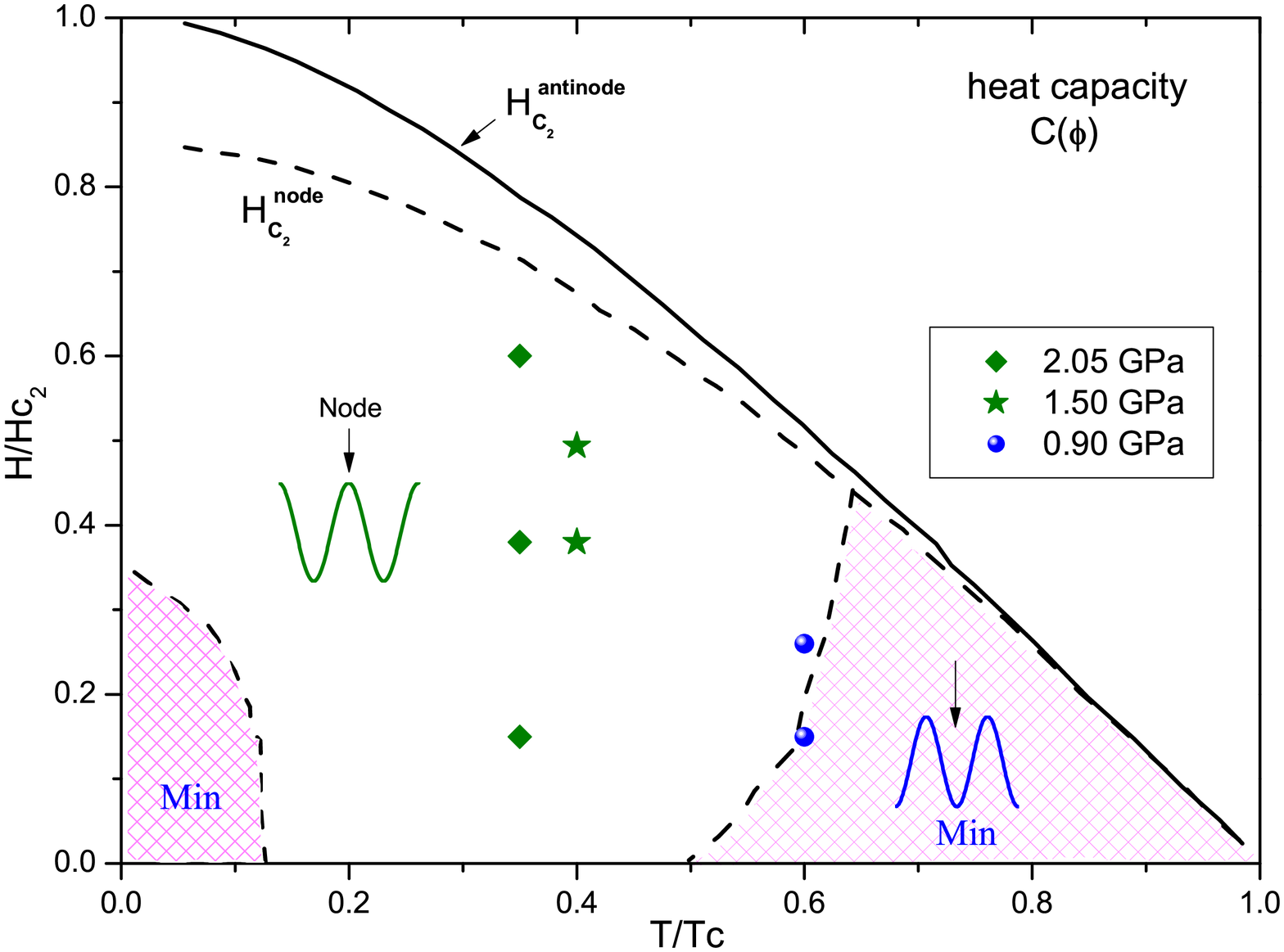}
\vspace{-17pt}
\caption{\label{fig:CeIrIn5H-Tphase} (color online). The calculated four-fold oscillation of  field-rotation heat capacity in a normalized H-T phase diagram, taken from Ref. \onlinecite{IVekhter08Fermisurface}. Our data points are plotted in the H-T phase diagram for CeIrIn$_5$ at 2.05 (diamonds), 1.50 (stars) and 0.90 GPa (circles).}
\vspace{-8pt}
\end{figure}

This interpretation is supported by results plotted in Figs. \ref{fig:CeIrIn5FieldRotation} (e)-(g) that show the field-rotation heat capacity C$_{ac}$ of CeIrIn$_5$ at 0.90 GPa and 0.3 K as a function of field angle $\phi$ for H=2.0, 3.5 and 5.0 kG, respectively. In contrast to the data at 2.05 GPa, four-fold oscillations at 2.0 and 3.5 kG are inverted and show minima along [110]. Figure \ref{fig:CeIrIn5FieldRotation} (h) is a plot of the normalized four-fold oscillation amplitude C$_4$/$\Delta$C(H) as a function of normalized field H/H$_c$ at 0.3 K for CeIrIn$_5$ at 2.05, 1.50 and 0.90 GPa, where $\Delta$C(H)=C(H)-C(0) is the change of heat capacity in magnetic field. With similar amplitudes (3 - 6\%), the four-fold oscillations have the same sign  at 2.05 and 1.50 GPa but a different sign at 0.90 GPa. We summarize all of our results at different pressures and magnetic fields and compare them to the theoretical expectations \cite{IVekhter08Fermisurface} in Fig. \ref{fig:CeIrIn5H-Tphase} . This comparison assumes the similarity between CeCoIn$_5$ and CeIrIn$_5$ mentioned above. While the measurements at 2.05 and 1.50 GPa are located in the intermediate region, the points at 0.90 GPa and T/T$_c\sim$ 0.6 are near the boundary of the low H/H$_c$, high T/T$_c$ shaded region where the four-fold oscillation swtiches its sign again from that expected in the intermediate region and exhibit minima for field in the nodal directions. We note that the detailed division of different regions depends on the exact shape of the Fermi surface, and the general trend of sign change from intermediate to high temperature ranges should survive \cite{IVekhter08Fermisurface}.   Consequently, oscillation minima along [110] for CeIrIn$_5$ at 0.90 GPa also are consistent with d$_{x^2-y^2}$ OP symmetry.

An interesting point is that we observe an in-plane upper critical field anisotropy, with H$_{c2}$ in the [110] direction always smaller than along [100] for all pressures.On the basis of  model calculations in Ref. \onlinecite{IVekhter07heatcapacity}, this upper critical field anisotropy between nodal and antinodal direction appears naturally as a result of the d-wave gap symmetry with H$^{node}_{c2}<$H$^{antinode}_{c2}$. We also stress that the in-plane H$_{c2}$ anisotropy cannot be the origin of the observed four-fold heat capacity oscillation because the oscillation change its sign between 2.05 and 0.90 GPa but the in-plane H$_{c2}$ anisotropy remains the same. We thus conclude that the line nodes are in [110] directions and the gap symmetry of CeIrIn$_5$ stays d$_{x^2-y^2}$ across SCII.

Because of proximity to AFM order, it is natural to consider that the OP symmetry in SCI is d$_{x^2-y^2}$ as it is in CeCoIn$_5$ or CeRhIn$_5$ under pressure, which also are near AFM order. On the other hand, our field-rotation heat capacity results also favor d$_{x^2-y^2}$ gap symmetry deep inside SCII, up to 2.05 GPa. In this regard, recent transport measurements observe non-Fermi liquid behaviors and claim that AFM fluctuations survive deep into SCII \cite{YNakajima08CeIrIn5}. It is then an interesting question why the superconducting phase apparently is divided into two regions but with the same gap symmetry. One possible scenario is that there is some hidden competing order that destroys bulk superconductivity in CeIrIn$_5$ near ambient pressure and produces a minimum in the bulk superconducting transition temperature. More studies are required to clarify the origin of two superconducting domes in the CeIrIn$_5$ system.

We are grateful to valuable discussions with Tanmoy Das, M. J. Graf and I. Vekhter. Work at Los Alamos was performed under the auspices of the U.S. Department of Energy, Division of Materials Science and Enginneering and supported in part by the Los Alamos LDRD program. TP was supported by the National Research Foundation (NRF) grant (2009-0075786) funded by Korea government (MEST).

\bibliographystyle {apsrev} 

\end{document}